\documentclass[pra,twocolumn,10pt,amsmath, superscriptaddress,notitlepage,showpacs]{revtex4-1}
\usepackage{amsmath,amsfonts,amssymb,amstext,amscd,amsthm,dsfont,bbm,hyperref,natbib}
\usepackage{physics}
\usepackage{color}
\usepackage{soul,xcolor}
\usepackage{gensymb}
\hypersetup{
    colorlinks = true,
    citecolor=blue,
    linkcolor = red,
    anchorcolor = red,
    citecolor = blue,
    filecolor = red,
    pagecolor = red,
    urlcolor = red,
    }
\usepackage[normalem]{ulem}
\usepackage{graphicx}
\usepackage{bbold}
\usepackage{braket}
\usepackage{placeins}

\renewcommand{\min}{{\mathrm min}}

\begin{document}
\title{Convex combinations of CP-divisible Pauli channels that are not semigroups}
	\author{Vinayak Jagadish}
	\affiliation{Quantum Research Group, School  of Chemistry and Physics,
		University of KwaZulu-Natal, Durban 4001, South Africa}\affiliation{ National
		Institute  for Theoretical  Physics  (NITheP), KwaZulu-Natal,  South
		Africa}
			\author{R. Srikanth}
	\affiliation{Poornaprajna Institute of Scientific Research,
		Bangalore- 560 080, India}
	\author{Francesco Petruccione}
	\affiliation{Quantum Research Group, School  of Chemistry and Physics,
		University of KwaZulu-Natal, Durban 4001, South Africa}\affiliation{ National
		Institute  for Theoretical  Physics  (NITheP), KwaZulu-Natal,  South
		Africa}

\begin{abstract} 
	We study the memory property of the channels obtained by convex combinations of Markovian channels that are not necessarily quantum dynamical semigroups (QDSs). In particular, we characterize the geometry of the region of (non-)Markovian channels obtained by the convex combination of the three Pauli channels, as a function of deviation from the semigroup form in a family of channels. The regions are highly convex, and interestingly, the measure of the non-Markovian region shrinks with greater deviation from the QDS structure for the considered family, underscoring the counterintuitive nature of (non-)Markovianity under channel mixing. 
\end{abstract}
\maketitle  
\section{Introduction}
Non-Markovian dynamics of open quantum systems~\cite{petruccione} is an active area of research, throwing new challenges and surprises~\cite{vacchini_markovianity_2011,breuer_colloquium:_2016,li_concepts_2017,de_vega_dynamics_2017,li_non-markovian_2019,li_non-markovian_2020}. The finite-time dynamics of open quantum systems  are described by time-dependant completely positive trace preserving (CPTP) maps, usually referred to as quantum channels~\cite{sudarshan_stochastic_1961, Quanta77}. Quantum non-Markovianity, unlike its classical counterpart does not have a unique definition and mathematical characterization. The two widely used approaches to study quantum non-Markovianity,  are based  on a deviation from CP-divisibility criterion~\cite{rivas_entanglement_2010, hall2010} and on the distinguishability  of  states~\cite{breuer_measure_2009}.  It is of interest to note that earlier non-Markovianity had been identified with  the quantum dynamical semigroup (QDS) structure. This was motivated by the fact that it can reasonably be considered as a quantum extension of the Chapman-Kolmogorov equation in the context of classical Markovianity, and that it corresponds to a weak system-environment coupling \cite{breuer_colloquium:_2016,chruscinski_non-markovian_2010}. More recently, \cite{shrikant2019concept} has argued that any deviation from the QDS form encodes a weak kind of memory in that the intermediate map lacks form-invariance.

Convex combinations of quantum channels have been actively studied recently \cite{wolf_assessing_2008,chruscinski_non-markovianity_2015,wudarski2016,megier_eternal_2017,breuer_mixing-induced_2018,jagadish_convex_2020}.  In the last cited, we considered the problem of mixing three Pauli channels, each assumed to be a QDS, and obtained a quantitative measure of the resulting set of Markovian and non-Markovian (CP-indivisible) channels. The present work leverages the technical content of \cite{jagadish_convex_2020} to address a qualitatively new question: whether or not convex combinations of channels that deviate more from Markovian semigroups produces more non-markovianity. Prima facie, the above observations suggest that if one were to mix channels that deviate from QDS, and which are thus less Markovian in the sense mentioned above, then this would correspondingly result in a larger measure of non-Markovian channels over different combinations. Surprisingly, this turns out not to be the case, as we show here. 
 
  The paper is organized as follows.  We present the preliminaries and discuss the convex combination of the three Markovian Pauli channels which are not QDSs. We then characterize the geometry of the (non-)Markovian region obtained by mixing, and evaluate its measure. Further, the behaviour of the regions as a function of deviation of the mixing channels from the QDS form is discussed. 
\section{Convex Combinations of Pauli Channels}
\label{convex}

We consider arbitrary convex combinations of the three Pauli channels. They are defined as
\begin{eqnarray}
 \Phi_x^q(\rho)&=&(1-q)\rho + q\sigma_x\rho \sigma_x,\nonumber\\
\Phi_y^q(\rho)&=&(1-q)\rho + q\sigma_y\rho \sigma_y \nonumber\\
\Phi_z^q(\rho)&=&(1-q)\rho + q\sigma_z\rho \sigma_z,
 \label{paulichann}
\end{eqnarray}

 where $\sigma_i$'s are the Pauli matrices. The general form of the three-way mixing is described by
 \begin{equation} \tilde{\Phi}_\ast(q) = x\Phi_x^q + y\Phi_y^q+ z\Phi_z^q,
 \label{threechanneleq}
 \end{equation}
 with $x, y, z \ge 0$ and $x+y+z=1$ and $q$ is a decoherence parameter, which in general is time-dependent.
The set of all channels of the form Eq. (\ref{threechanneleq}) constitutes the \textit{Pauli simplex}, whose vertices are the Pauli channels assumed to be described by the same parameter $q$ \cite{jagadish_convex_2020}.

We now choose $q$ from the family with the functional form
 \begin{equation}
q =  \frac{1-\mathrm{exp}(-rt)}{n},
\label{eq:q}
 \end{equation} 
 with $n$ being any positive real number greater than or equal to 2, and $r$ being a constant. For the channel, $\Phi_z$, the corresponding time-local generator (defined by $\dot{\Phi} = L(t)\Phi$) reads
\begin{eqnarray}
\label{megen}
L(t) \rho &=& \gamma^{(n)} (t) (\sigma_z\rho\sigma_z-\rho),\nonumber\\
\gamma^{(n)}(t) &=& \frac{r}{(n-2)e^{r t}+2},
\end{eqnarray}
with the time-dependence of the decay rate, $\gamma^{(n)}(t) $ showing that $L(t)$ generate a semigroup only for $n=2$, where $\gamma^{(2)} = \frac{r}{2}$, being time-independent.
 The reason for choosing the particular form of $q$ as in Eq. (\ref{eq:q}) is to make a comparison with QDS. It can be easily seen that the only choice for a Pauli channel Eq. (\ref{paulichann}) to be a semigroup is the one corresponding to $q =  \frac{1-\mathrm{exp}(-rt)}{2}$.

Now, the time-local generator for the channel $\tilde{\Phi}_\ast(q)$, Eq. (\ref{threechanneleq}) follows to be of the form
 \begin{equation}
\label{megen}
L(t) \rho = \sum_{k = x,y,z}\gamma_{k}(t)(\sigma_k\rho\sigma_k-\rho),
\end{equation} with the decay rates being 
\begin{eqnarray}
\gamma_x &=& \left(\frac{1-y}{1-2 (1-y)q}+\frac{1-z}{1-2 (1-z)q}-\frac{1-x}{1-2 (1-x) q}\right)\frac{\dot{q}}{2}\nonumber\\
\gamma_y &=&\left(\frac{1-x}{1-2 (1-x)q}+\frac{1-z}{1-2 (1-z)q}-\frac{1-y}{1-2 (1-y) q}\right)\frac{\dot{q}}{2}\nonumber\\
\gamma_z &=&\left(\frac{1-x}{1-2 (1-x)q}+\frac{1-y}{1-2 (1-y)q}-\frac{1-z}{1-2 (1-z) q}\right)\frac{\dot{q}}{2}. 
\label{ratesdecaythree}
\end{eqnarray}
The study of these rates is largely simplified because the summands that make them up have the same functional form. This can be exploited to quantify the measure of the region of non-Markovian channels.

\section{Geometry and Measure of (non-)Markovian regions}
\label{measure}

Given a convex mixture of the three Pauli channels, we are now in a position to discuss the geometry of the Markovian and non-Markovian regions in the parameter space of $(x,y)$ and to analytically evaluate the corresponding measure of the regions. Here it is worth pointing out that there have been a number of criteria and corresponding measures that have been proposed to witness and quantify non-Markovianity~\cite{breuer_colloquium:_2016,li_concepts_2017,rivasreview}. The two major approaches are based on CP-divisibility~\cite{rivas_entanglement_2010, hall2010}, and on the distinguishability  of  states~\cite{breuer_measure_2009}. 
\begin{itemize}
\item RHP divisibility criterion~\cite{rivas_entanglement_2010}: A quantum channel is Markovian if it is CP-divisible at all instants of time. Any deviation from CP-divisibility is an indicator of non-Markovianity according to RHP criterion. 
\item HCLA Criterion~\cite{hall2010}: A dynamics generated by a master equation of the form Eq. (\ref{megen}) is  Markovian if and only if all the decay rates $\gamma_{k}(t)$ are non-negative. So, if any one of the decay rates turn negative at any instant of time, the channel is non-Markovian. This can be shown to be equivalent to the RHP criterion. In what follows, the characterization of non-Markovianity is therefore done by analyzing the decay rates in the time-local master equation corresponding to the channels.
\item BLP distinguishability or information backflow criterion~\cite{breuer_measure_2009}: A quantum channel $\mathcal{E}(t)$ is Markovian if it does not enhance the distinguishability of two initial states $\rho_A$ and $\rho_B$, i.e., if $\Vert\mathcal{E}(t)(\rho_A) - \mathcal{E}(t)(\rho_B)\Vert \le \Vert\mathcal{E}(0)(\rho_A) - \mathcal{E}(0)(\rho_B)\vert\vert$, where $\Vert \cdot \Vert$ denotes the trace distance. For qubits, this is known to be equivalent to P-divisibility~\cite{chruscinski_non-markovianity_2015}, and thus provides a stronger criterion of non-Markovianity than CP-indivisibility.

From Eq. (\ref{ratesdecaythree}), we can see that the decay rate expressions have the form 
\begin{eqnarray}
\gamma_x(x,y,z) &=& -f(x,p)+f(y,p)+f(z,p) \nonumber \\
\gamma_y(x,y,z)  &=& f(x,p)-f(y,p)+f(z,p) \nonumber \\
\gamma_z(x,y,z)  &=& f(x,p)+f(y,p)-f(z,p),
\label{eq:3form}
\end{eqnarray} 
where $f(\alpha,p)\ge0$ for all $p \in [0,\frac{1}{n})$ and $\alpha \in \{x,y,z\}$. An immediate consequence is that the sum $\gamma_{a} + \gamma_{b} , a,b = x,y,z , a \neq b$ is always positive, even though an individual rate may be negative. For example $\gamma_y + \gamma_z = 2f(x,p) \ge 0$.  This implies that the dynamics obtained by convex combination is P-divisible and hence Markovian for channels on a qubit, according to the BLP distinguishability criterion \cite{chruscinski_non-markovianity_2015}. 
\end{itemize}
The resultant channels, Eq. (\ref{threechanneleq}) obtained by convex combinations of Pauli channels which are not semigroups are always P-Divisible, and hence Markovian according to the BLP criterion. We therefore identify quantum non-Markovianity with CP-indivisibility based on the analysis of the decay rates, Eq. (\ref{ratesdecaythree}) in the time-local master equation corresponding to the channels.

It can be shown that the structure of Eqs. (\ref{ratesdecaythree}) guarantees that if a given rate (say) $\gamma_y(x,y,z=1-x-y,q)$ turns negative at $q=q_0 \le \frac{1}{n}$, then it remains negative throughout the remaining range of $[q_0,\frac{1}{n}]$ \cite{jagadish_convex_2020}.  To find the set of all pairs $(x,y)$ such that $\gamma_y(x,y,q)\le0$ at $q=\frac{1}{n}$, we solve the equation $\gamma_y(x,y,\frac{1}{n}) = 0$. The result is a constraint on the pairs $(x,y)$, which can be represented by expressing $x$ in terms of $y$:

\begin{equation}
x_\pm(y) \equiv \frac{1}{2} 
 \left(\pm\frac{g(n,y)}{y+(n-1)}-y+1\right),
\label{eq:xmaxmin}
\end{equation}
where
\begin{align}
g(n,y) &= \left[(-n+y+1) (n+y-1) \left(\beta_n^+-y\right)\left(\beta_n^--y\right)\right]^{\frac{1}{2}},\nonumber\\
\beta_{n}^{\pm} & = \pm\sqrt{n^2+1}-n.
\end{align} 
The values $x_\pm(y)$ are real only in the range $y \in [0, \beta_n^+]$.

Further, the form of Eq. (\ref{eq:xmaxmin}) means that for any given $y$ in the above allowed range, the values $x \in (x_-(y), x_+(y))$ yield $\gamma_y<0$, and those outside, i.e., the values $x \in [0, x_-(y)] \cup [x_+(y),1]$, yield $\gamma_y\ge0$. Thus, we determine the region $\mathcal{R}_y$ as corresponding to these points $(x, y)$ which yield a negative $\gamma_y$:
\begin{equation}
|\mathcal{R}_y| = 2  \int_{y=0}^{\beta_{n}^+} \left(x_+(y) - x_-(y)\right)dy.
\label{eq:mu1}
\end{equation} 
The pre-factor 2 comes from the fact that the space of $(x,y)$ does not have area 1 but instead must be normalized to $\int_{x=0}^1 \int_{y=0}^{1-x} dx ~dy = \frac{1}{2}$. The form of the rates Eq. (\ref{ratesdecaythree}) is such that at most only one of the three rates can be negative \cite{jagadish_convex_2020}. This means that regions $\mathcal{R}_x, \mathcal{R}_y$ and $\mathcal{R}_z$, respectively, of points $(x,y,z)$ where $\gamma_x, \gamma_y$ and $\gamma_z$, can assume negative values within the time range $q \in [0, \frac{1}{n}]$, is non-overlapping. Therefore, the measure, $|\overline{\mathcal{M}}|$ of the set  of all non-Markovian channels in the Pauli simplex $\mathcal{P}$,  is simply $|\overline{\mathcal{M}}| = 3|\mathcal{R}_y|$.  

\begin{figure}[t!]
	\includegraphics[width=80mm]{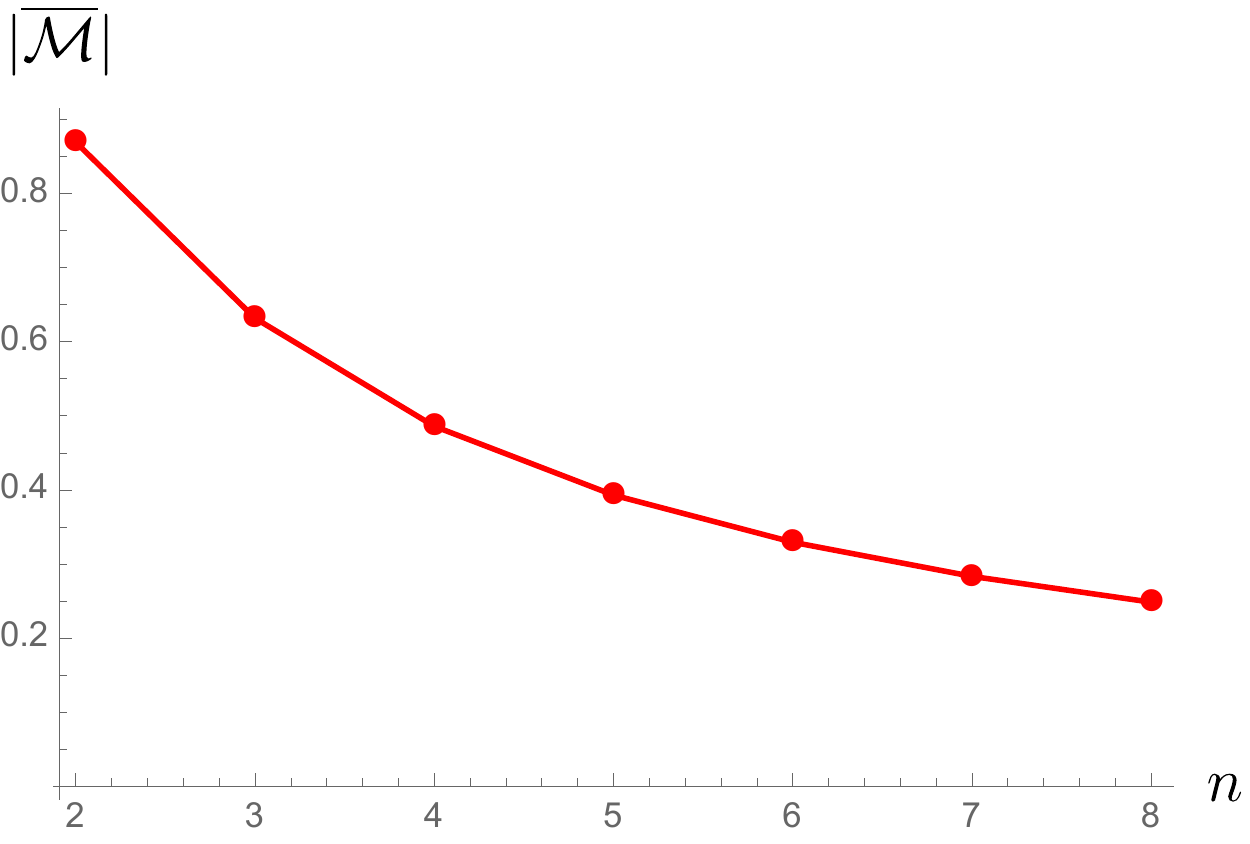}
	\caption{(Color online) Plot of the measure of non-Markovian channels in the Pauli simplex, $|\overline{\mathcal{M}}|$ with varying $n$. One finds that $|\overline{\mathcal{M}}|$ decreases with increasing $n$. The case $n=2$ corresponds to QDS.}
	\label{fig:measure}
\end{figure}

A plot of the measure $|\overline{\mathcal{M}}|$  of non-Markovian regions with varying $n$  is shown in Fig. \ref{fig:measure}. It shows that as the mixing channels move to a greater degree $n$ away from QDS ($n=2$), somewhat counter-intuitively, the fraction of non-Markovianity in the corresponding Pauli simplex falls.  
	\begin{figure}[t!]
	\includegraphics[width=80mm]{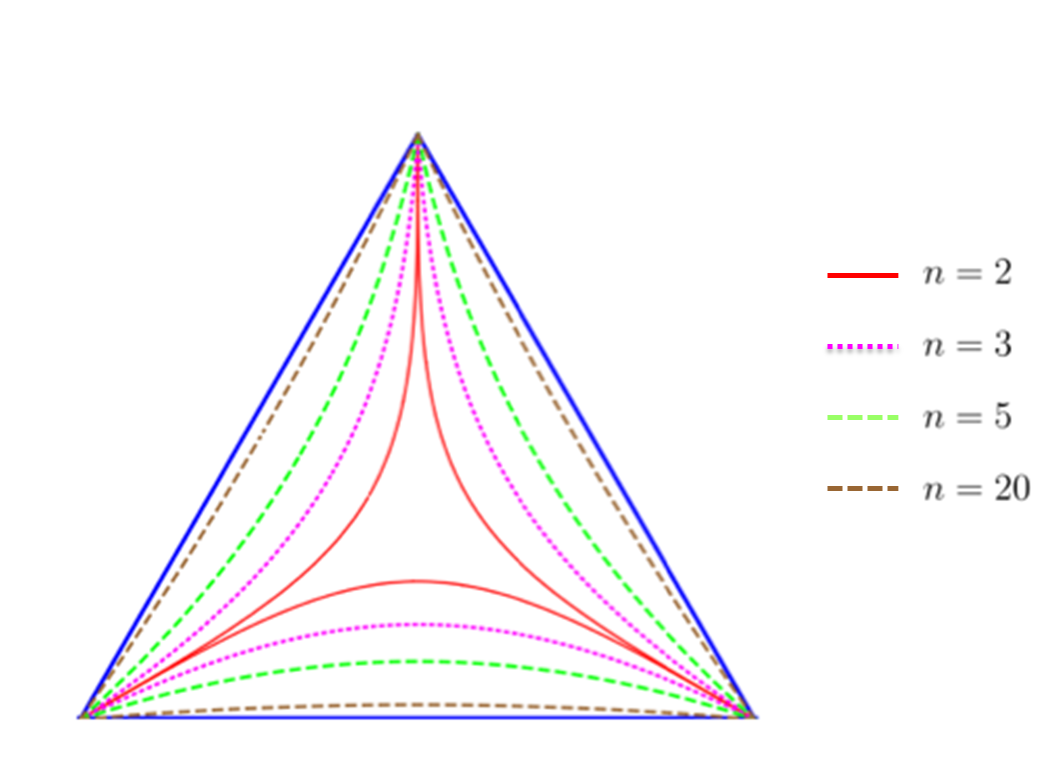}
	\caption{(Color online) The outermost triangle (in blue) represents the Pauli simplex for a given functional form $q(n)$, with the vertices representing the three Pauli channels. The squeezed triangles represent the Markovian regions $\mathcal{M}_n$ of Markovianity for different degrees $n$ of deviation from the QDS value of $n=2$. We note that $\mathcal{M}_n \subset \mathcal{M}_{n^\prime}$ if and only if $n < n^\prime$. }
	\label{fig:pauli}
	\end{figure}
The natural diagrammatic depiction of the Pauli simplex as per our above analysis is in the $(x,y)$ representation, or analogously in the corresponding $(x,z)$ or $(y,z)$ representation. This is a right angle triangle (bordered by $y=1-x$). To go to a ``Pauli neutral'' representation, we require the  linear transformation that maps a right angle triangle with vertices $\{(0,0), (0,1), (1,0)\}$ to an equilateral triangle. This is given by the matrix
$M \equiv k\left(
\begin{array}{cc}
	2 & 1 \\
	0 & \sqrt{3} \\
\end{array}
\right)$, where $k$ is a constant set to $\sqrt{\frac{1}{2\sqrt 3}}$ to ensure that the transformation is area preserving (i.e., det($M$)=1). The Pauli simplex in this representation corresponds to the equilateral triangle $\{(0,0), (\frac{1}{2}, \frac{\sqrt3}{2}), (1,0)\}$. The Markovian squeezed triangular regions $\mathcal{M}_n$ are mapped correspondingly, as depicted in Fig. \ref{fig:pauli}. Here, the equilateral triangle corresponds to a Pauli simplex for any $n$ with the corresponding Pauli channels of the type Eq. (\ref{eq:q}). 

\begin{figure}[t!]
	\includegraphics[width=85mm]{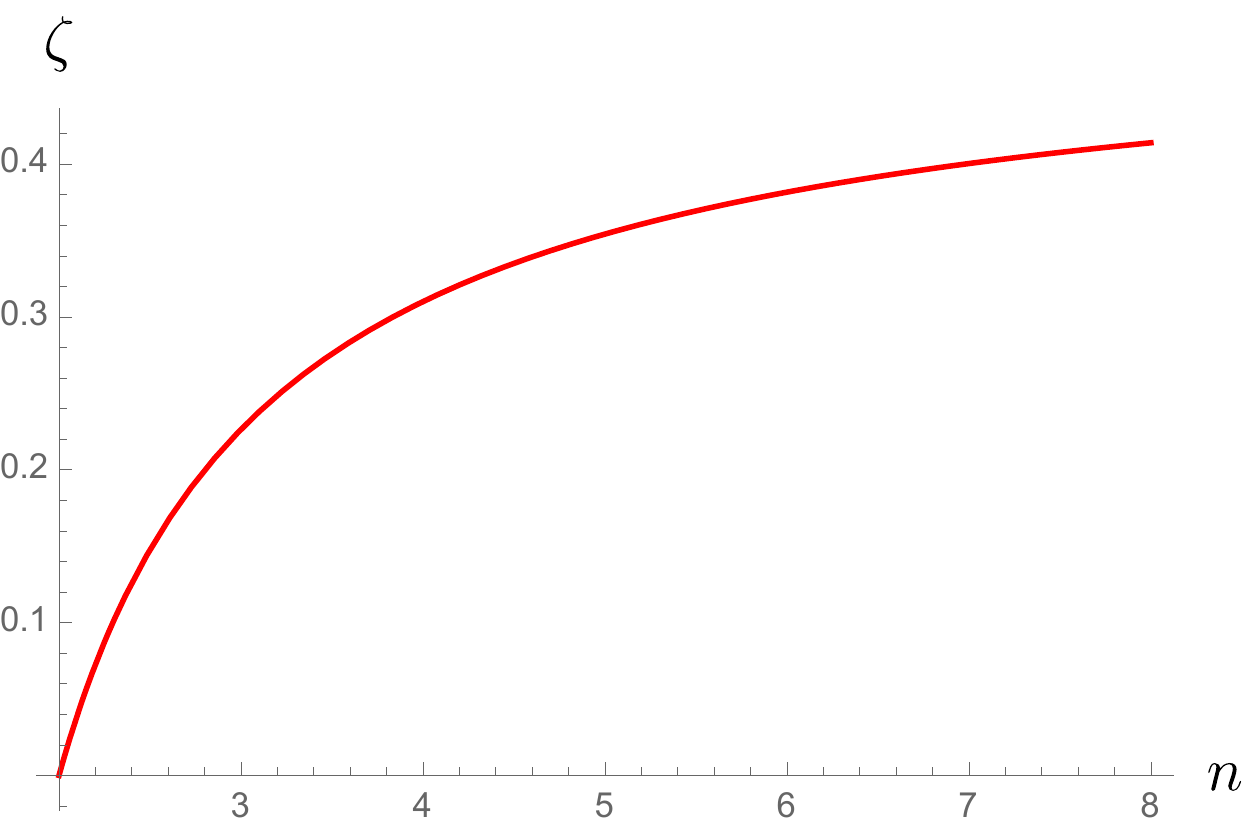}
	\caption{(Color online) Plot of the measure of non-Markovianity, $\zeta$ with varying $n$, evaluated using Eq. (\ref{qdsnm1}). One finds that $\zeta$ increases with increasing $n$.}
	\label{fig:qdsnm}
\end{figure}

Fig. \ref{fig:pauli} shows that as the degree $n$ of deviation from QDS form increases, the Markovian regions corresponding to a larger deviation contain those of a smaller deviation in the Pauli simplex, i.e. $\mathcal{M}_n \subset \mathcal{M}_{n^\prime}$ if and only if $n < n^\prime$. Certain points of similarity with the QDS case may be worth noting: in the case of two-channel mixing, which corresponds to any edge of the Pauli simplex, note that the result is the same as the QDS case: namely, any finite mixing leads to non-Markovianity. 
One way to understand this surprising result is to note, in view of Eq. (\ref{eq:q}), that a larger $n$ corresponds to channels that decohere to a lesser degree. From that perspective, the mixing can be expected to produce a larger region corresponding to Markovian channels.  A recent approach to non-Markovianity identifies a deviation from the QDS form as a weak form of memory, in that it corresponds to the loss of a strong concept of memorylessness called temporal self-similarity of the quantum channel \cite{shrikant2019concept}. Accordingly, non-Markovianity in a weaker sense may be geometrically quantified by the minimum distance of an evolution from the QDS form (evaluated at the level of generators) and given by
\begin{equation}
\zeta = \min_{L^\ast} \frac{1}{T} \int_0^T \Vert \hat{L}(t) -  \hat{L}^\ast\Vert dt,
\label{qdsnm}
\end{equation}
where $\hat{L}$ is the generator applied to one half of a singlet state. For the present case, Eq. (\ref{qdsnm}) evaluates to
\begin{equation}
\zeta = \int_0^1 (\gamma^{(n)}(t) - \gamma^{(2)})  dt.
\label{qdsnm1}
\end{equation}
Setting the constant $r$ to be unity, a plot of $\zeta$ for various $n$ (Fig. \ref{fig:qdsnm}) shows that as $n$ increases the measure increases, showing greater non-Markovianity from the perspective of QDS as Markovian. Note that the three Pauli channels for the entire considered range of $n$ are Markovian according to the above-mentioned stronger criteria of non-Markovianity, such as those based on divisibility or distinguishability, which would thus make these stronger criteria unsuitable to highlight the element of surprise about Figure \ref{fig:pauli}.

Finally, as in the QDS case, for any $n$ neither the set of Markovian nor that of non-Markovian channels in the Pauli simplex is convex. In Fig. \ref{fig:pauli}, line segments or triangles connecting the ``horns'' of the squeezed triangle give us infinite number of examples of non-Markovian channels obtained by mixing Markovian channels. On the other hand, line segments or triangles linking the convex regions $\mathcal{R}_j$ outside the squeezed triangles give an infinite number of examples of Markovian channels obtained by mixing non-Markovian ones.

\begin{figure}[t!]
	\includegraphics[width=85mm]{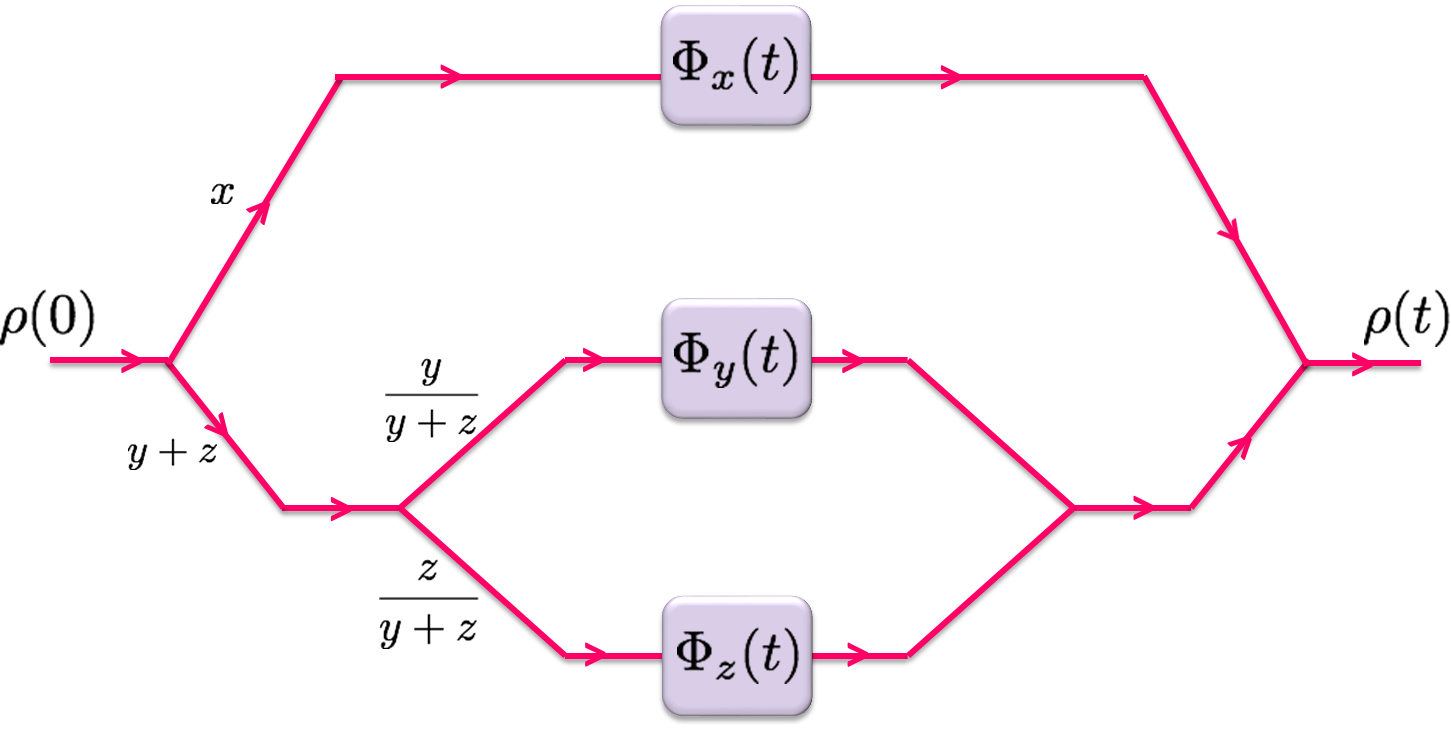}
	\caption{(Color online) Proposed optical setup for the implementation of convex combinations of the three Markovian  channels}
	\label{fig:expsetup}
\end{figure}

Fig. \ref{fig:expsetup} shows a suggested optical setup for the implementation of convex combinations of the three Markovian  channels. First the light on one arm is split using a biased beamsplitter with bias $x$ on one side and $y+z$ on the other. On the $x$ arm, the channel $\Phi_x(t)$ is applied through suitable optical elements. Now the other arm is subjected to a second biased beamsplitter with biases $y/(y+z)$ and $z/(y+z)$. On one arm $\Phi_y(t)$ and on the other $\Phi_z(t)$ is applied. They are then recombined (lossily) into a single beam to produce the final beam which is recombined with beam $x$.

The above experiment is well within current quantum technology, and could be implemented by a parametric downconversion setup. In practice, it may be tedious to produce the mixed channel $\tilde{\Phi}_\ast(q)$ for a large number of values of the triples $(x,y,z)$, and thus a selection of these triples may be chosen to ensure a reasonable sampling of the Pauli simplex and verification of the pattern of Fig. \ref{fig:pauli}.

\section{ Discussions and Conclusions}
\label{conclusion}
We have studied the convex combination of Markovian Pauli non-QDS channels. The Pauli simplex obtained by the convex combination of the three Pauli channels is characterized and the measure of the associated non-Markovian regions is evaluated analytically.  For the family of channels parametrized by mixing fraction Eq. (\ref{eq:q}), the measure of the non-Markovian region in the Pauli simplex is found to decrease for mixing of channels that deviate more from the QDS structure. In other words, mixing time-dependent Markovian channels results in the production of ``more'' Markovian channels in comparison to mixing Markovian semigroups.

From Eq. (\ref{ratesdecaythree}), it follows that the functional form of the mixing fraction $q=q(t)$  determines the instant $q_0$ at which a given channel $\tilde{\Phi}_\ast(t)$ in Eq. (\ref{threechanneleq}) turns non-Markovian. However, we note from the form Eq. (\ref{eq:xmaxmin}) that the non-Markovian regions don't depend on the functional form but only the value $\frac{1}{n}$ that $q(t)$ asymptotes to.  This means, for example, that, as far as the measure of (non)-Markovian channels is concerned, for any fixed $n$, all channels corresponding to $q = [1- \exp(-rt^{m_1})]^{m_2}/n$, with $m_j$ being a real number greater than 1, are mutually equivalent. 

In~\cite{puchala_2019}, it was shown that the set of dynamical maps accessible through continuous semigroups is unitarily equivalent to a unistochastic channel. It would be worth investigating as to how it could be extended to channels which are not Markovian semigroups, based on the results that we have obtained in this paper. 
With the recent advances in simulating open quantum systems and quantum non-Markovianity by optical setups~\cite{obando2020,passos19}, we anticipate that our results can be implemented experimentally. 

Finally, it may be noted that semi-Markovian maps~\cite{breuersemimarkov,chruscinski_generalized_2017} which are CP-indivisible may be considered as weakly non-Markovian in the sense of Ref. \cite{shrikant2019concept} (i.e., deviating from QDS), and thus the mixing of semi-Markovian maps is expected to bring out similar features as reported here.

\section{Acknowledgments}  The work of V.J. and F.P. is based upon research supported by the South African Research Chair Initiative of the Department of Science and Innovation and National Research Foundation (NRF) (Grant UID: 64812). R.S.   thanks the
Department of Science and Technology (DST), India, Grant No.: MTR/2019/001516.

\bibliographystyle{apsrev4-1}
\bibliography{NonQDS.bib}
\end{document}